# From Readership to Usership and Education, Entertainment, Consumption to Valuation: Embodiment and Aesthetic Experience in Literature-based MR Presence

Stéphanie Bertrand[1], Martha Vassiliadi[2], Paul Zikas[3,5], Efstratios Geronikolakis[3,5], George Papagiannakis[3,4,5]

[1] School of Architecture, Faculty of Engineering Aristotle University of Thessaloniki, Thessaloniki, Greece
bertrandstephanie@gmail.com

[2] Department of Philology, School of Philosophy, Aristotle University of Thessaloniki, Thessaloniki, Greece
marthatv@lit.auth.gr

[3] Foundation for Research and Technology Hellas, 100 N. Plastira Street, 70013 Heraklion, Greece

[4] Computer Science Department, University of Crete, Voutes Campus, 70013 Heraklion, Greece

[5] ORamaVR, 100 N. Plastira Street, 70013 Heraklion, Greece
{paul, stratos, george.papagiannakis}@oramavr.com

**Abstract**

This chapter will extend its preliminary scope by examining how literary transportation further amplifies presence and affects user response vis-à-vis virtual heritage by focusing on embodiment and aesthetic experience. To do so, it will draw on recent findings emerging from the fields of applied psychology, neuroaesthetics and cognitive literary studies; and consider a case study advancing the use of literary travel narratives in the design of DCH applications for Antiquities – in this case the well-known ancient Greek monument of Acropolis. Subsequently, the chapter will discuss how Literary-based MR Presence shifts public reception from an education-entertainment-touristic consumption paradigm to a response predicated on valuation. It will show that this type of public engagement is more closely aligned both with MR applications' default mode of usership, and with newly emerging conceptions of a user-centered museum (e.g., the Museum 3.0), thus providing a Virtual Museum model expressly suited to cultural heritage.

**Introduction**

Since 2014, with the exponential adoption of consumer Virtual Reality (VR) technology, cultural institutions have increasingly turned towards Mixed Reality (MR) applications to expand and democratize public access to heritage. However, recent findings have shown that existing MR intangible and tangible digital cultural heritage applications by and large fail to adequately engage audiences beyond an initial fascination with the immersive 3D visualization on account of either misguided storytelling, or non-compelling, non-engaging narratives (Vassiliadi et al. 2018). In response, the concept of Literature-based MR Presence was recently introduced to address the 'content-based' shortcomings of modern MR intangible and tangible digital heritage storytelling applications (Vassiliadi et al. 2018). This notion was first established on the basis of literary myth's potential to enhance presence in Digital Cultural Heritage immersive applications owing to its multi-temporal and multi-cultural features. As argued, these features are coextensive with Mixed Reality applications' multimodal functions, and moreover have a beneficial impact on the viewer. In effect, they amplify presence not only by providing a more engaging narrative than the straightforward delivery of informational content; but also through the phenomenon of literary transportation.

## 1. Presence and Literary Transportation

As outlined above, the initial case for Literary-based MR Presence was first established by focusing on: a) the potency of storytelling vis-a-vis didactic content (i.e., the attention-grabbing capability of captivating narratives); b) the multi-temporal and multi-cultural aspects of literary myths (i.e., fiction's suitability to XR's multimodal affordances); and c) the impact of literary transportation (Vassiliadi et al. 2018). Of the three characteristics originally ascribed to the concept, literary transportation is of particular importance to the idea of presence in virtual environments [VE], which is generally described as a subjective sensation of 'being there'. Indeed, as Schuemie et al. observe in their survey of presence in virtual environments: 'presence as discussed in literature related to immersive VR can most often be characterized by the concept of presence as transportation: people are usually considered "present" in an immersive VR when they report a sensation of being in the virtual world ("you are there")' (2001). The same immersive principles are applied in Augmented Reality (AR) environments through the introduction of True AR elements that blend the real with the virtual world (Geronikolakis et al. 2019). For this reason and on account of its correlation to the arguments put forward below, the term 'literary transportation' warrants a brief prefatory recap.

By definition, literary transportation is 'a convergent process, where all mental systems and capacities become focused on events occurring in the narrative' (Green & Brock, 2000), therefore offering a means to greatly enhance a user's engagement with a given simulation. It should be noted that this phenomenon also functions with respect to alternative understandings of presence in VE, such as Slater's Psi (whereby presence corresponds to the illusion that the environment exhibited in VR is actually taking place (Slater & Sanchez-Vives 2016)), as it simultaneously offsets the sensorimotor constraints of the VR system while increasing the credibility of the scenario. Moreover, literary transportation is associated with intense aesthetic involvement attributable to the activation of the brain's default mode network (Starr 2013, 59-63): a pattern of cognitive activity linked, among others, to internal mentation (Andrews-Hanna 2011), self-reflection and eudaimonia (personal growth and well-being) (Stark et al. 2018). As follows, the impact of literary transportation has far-reaching consequences for the user beyond heightened involvement and immersion in the MR environment – i.e., rendering the experience both more appealing and more credible. Literary transportation has also been empirically shown to increase empathy skills over time across fictional genres in contrast to non-fiction (Bal & Veltkamp 2013), thus presenting promising connexions with the emerging development of empathy training in VR (Bertrand et al. 2018; Schutte & Stilinović 2017).

With that knowledge in mind, beyond providing captivating narratives and triggering transportation, the use of literary texts in MR intangible and tangible digital heritage storytelling applications can also contribute to amplifying presence through embodiment. Specifically, literature has been shown to elicit embodied responses from readers through several means, including the use of multisensory imagery examined below. Before delving further into these processes, however, it is imperative to first consider embodiment's role in immersive digital applications so as to further contextualize and assess the impact of the proposed approach in relation to the existing scholarship.

## 2. Presence and Embodiment in Extended Reality (XR)

If 'presence is typically seen in academic research as the aim of virtual reality environments' (Pujol & Champion 2012), embodiment constitutes a vital yet still under investigated condition for its successful attainment. As enactivism, embodied cognition and situated cognition theories and empirical findings have shown over the past few decades – counter to prior cognitivist approaches and mind-body dualism – embodiment is an integral part of cognitive processing. With respect to VR, the chief implication of this understanding is that it bridges the relation of the self to its surrounding environment (i.e., external involvement and immersion) with the relation of the self to the body (i.e., internal sensations such as self-location and body ownership), which are fundamental to achieving a satisfactory level of presence in VE.

In terms of the existing literature, Biocca remarks, as early as 1997, that embodiment is expected to have a direct impact on different types of presence (Biocca 1997). Citing this research, Schubert et al. soon after propose the notion of 'embodied presence', which they attribute to meshed patterns of action primarily oriented towards navigation and interaction with objects in VE (1999). In a similar vein, ecological views of presence greatly indebted to Gibson have explored the bearing of situated affordances, perception-action coupling and 'ready-to-hand' tools for VR (Schuemie et al. 2001). However, broadly speaking, the field has produced only limited research into the potential impact of embodied cognition in the context of VE.

In a rare paper leveraging embodied cognition to advance 'a new set of relationships between dimensions of embodiment and forms of presence', Costa et al. have recently theorized that increased embodiment is correlated with higher levels of psycho-physiological responses to VE, and can therefore serve the purpose of improving the user's sense of presence (2013). Citing structural coupling's model of mutual affect between organism and environment, the authors affirm that: 'Practically speaking, we can say a user is embodied in a virtual reality environment (VRE) if changes in the VRE affect the user (emotionally, cognitively, and/or physiologically), and the user can affect the VRE (move objects, interact with others, etc.) (Costa et al. 2013).

Notwithstanding, as their notable paper makes plain, research into the interconnection between embodiment and presence in virtual environments to date remains primarily focused on sensory input, possible action and interaction in the VE, as well as on the user's relationship to their avatar; for instance, how the aforementioned innate senses of self-location and of body-ownership might be induced vis-à-vis an artificial body (Kilteni et al. 2012). When it comes to sensory input, as Costa et al. outline, empirical studies have concentrated on the impact of visual scale and dimensionality, the auditory effects of higher quality ambient and action driven sounds, and the haptic feedback offered by different controllers on the user (2013). In parallel to sensory engagement, the authors propose the notion of 'afforded embodiment' to capture the specific sense arising from avatar manipulation and customization, by relying on research demonstrating an increased sense of embodiment as a result of a) greater motor control (i.e., the degree to which the user can effectively control the avatar's movements), and b) psycho-social afforded embodiment (i.e., 'the degree to which the user can modify and/or manipulate their avatar to reflect or express their identity') (Costa et al. 2013). From a technical perspective, glGA (Papagiannakis et al. 2014) proposes a lightweight, open source, shader-based framework used in various graphics projects, among others,

VR exhibitions and rapid reconstruction of Cultural Heritage monuments in MR environments. A similar approach presents an AR system which relies purely on passive techniques to solve the real-time registration problems combining a VR component-based simulation framework with computer vision techniques to generate AR experiences (Vacchetti et al. 2004).

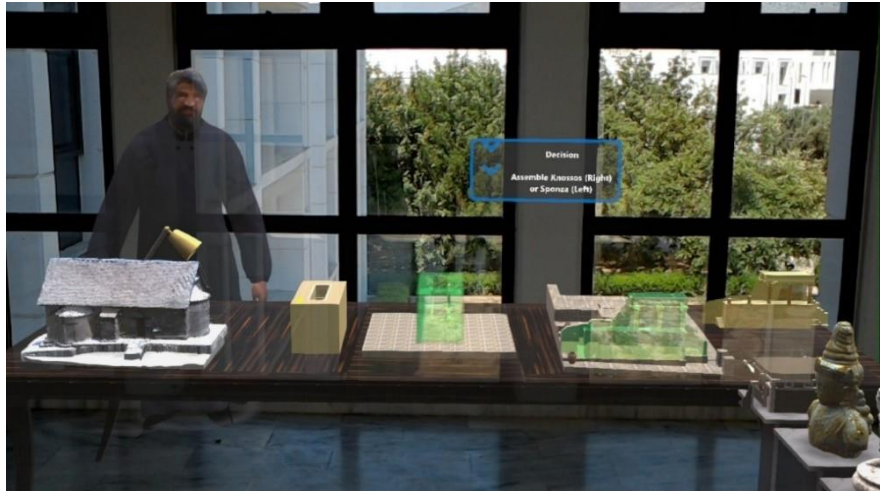

**Fig. 1** Illustrating an AR virtual museum. On the left, the 3D reconstructed and animated priest of Asinou church in Cyprus (Geronikolakis et at. 2019)

While sensory input, action & interaction and afforded embodiment are unquestionably vital to amplifying presence in MR applications, they fail to address how narrative content can also prime users and induce a heightened sense of embodiment in VE (though it should be noted here that Schubert et al. explicitly acknowledge dramatic content's measurable impact on presence in the context of screen-based 3D games (1999)). Examining this function of narrative content is particularly decisive in the case of virtual heritage projects, which necessitate added intellectual and emotional immersion to sustain 'cultural presence' (Pujol & Champion 2012, 98). As the following section will examine, new research in the fields of neuroaesthetics, cognitive cultural studies and cognitive semantics expressly focused on reader response to literary texts is beginning to provide new leads in this direction.

## 3. Literature, Embodiment and Mental Imagery

Departing from the first half of Costa et al.'s account according to which 'a user is embodied in a virtual reality environment (VRE) if changes in the VRE affect the user (emotionally, cognitively, and/or physiologically)' (2013), the case for Liter-ary-based MR Presence becomes clear-cut. To wit, the integration of literary texts into DCH applications' narrative content naturally heightens emotional response, and can reconcile a certain mediated external perception of heritage with a likely internal reaction. Hereof, as G. Gabrielle Starr relevantly points out in her examination of John Keats's "Ode on a Grecian Urn": 'while the urn can "express . . . more sweetly than . . . rhyme," poetry can evoke the visible surface of the urn as well as the internal response of a viewer—a sense of puzzlement at the urn's mysteries'. (Starr 2013, 12) But more than reflexively bridging outer engagement and mentation, the incorporation of literary texts into DCH applications can also play a seminal role in amplifying the user's sense of embodiment, and therefore presence in VE, through simulation (Gibbs Jr. 2017; Oatley 2011). Indeed, emerging research arising from the empirical turn in literary studies increasingly points towards embodied responses to poetry and fiction as a result of phenomena including multisensory imagery.

The latter is of particular significance to Literary-based MR Presence given poetry and fiction's distinctly vivid and sensuous portrayals. Beyond a certain baseline of literature-induced embodied simulation (whereby readers mirror protagonists' actions and emotions), these textual depictions have the effective ability to trigger powerful mental images: i.e., 'the subjective experience of sensation without corresponding sensory input' (Starr 2010, 276). As Kosslyn et al. observe, mental imagery occurs 'when a representation of the type created during the initial phases of perception is present but the stimulus is not actually being perceived; such representations preserve the perceptible properties of the stimulus and ultimately give rise to the subjective experience of perception' (2006, 4). In other words, when people experience mental imagery, the same areas of the brain that are involved in the perception of actual sensory input become active, and moreover operate in commensurate organizational patterns: for instance, auditory imagery is organized temporally, whereas visual and haptic images typically mirror spatial detail and relation (Starr 2013, 75). This phenomenon is observed across sensory modes: in fact, evidence points to many different types of mental imagery, including object-based imagery (e.g., shapes and colours), spatial imagery (e.g., of locations), auditory imagery and motor imagery: i.e., kinaesthetic and/or proprioceptive images (Moulton & Kosslyn 2009). In the context of VE, this suggests that it is not only direct sensory stimulation such as geographically encoded sound effects (see previous section) that can be mobilized towards increased embodiment. In addition, specific forms of narrative content, such as auditory images (in keeping with the example), can also contribute to amplifying this sense by activating many of the same areas of the brain in similar organizational patterns (Starr 2010).

Among the different types of sensory imagery, motor imagery has been put forward as the most paradigmatic case (Starr 2013, 81), in accordance with scientific evidence indicating a robust perception/action coupling when it comes to kinaesthetic and proprioceptive images. Indeed, neuroimaging studies have found significant overlap in the neural circuitry involved in action execution and in the obser-vation of another person's motions (e.g., seeing another person smile activates the same facial muscles at a sub-threshold level in the viewer), as well as in imagining both one's own actions and another person's actions (Decety & Jackson 2004). Reproducing images of somatic and motoric components is a means through which individuals recognize other people's emotions; and such reverse mapping has been linked to the architecture of empathy (Decety & Jackson 2004). Accordingly, motor imagery has been found to play a central role in theory-of-mind. Based on motion, that is, both the perception of biological motion and static images of motion, individuals infer other people's mental states: e.g., attribute intention – a crucial survival skill that has, consistently, been observed as early as preverbal infancy (Blackemore & Decety 2001).

It is no wonder then, as Starr remarks, that: 'Multisensory imagery, especially the multisensory imagery of motion, is centrally important to a variety of aesthetic pleasures in part because it gives us access not to the "real" complexity of experience but to certain powerfully connected aspects of the ways our minds internally represent experiences and objects.' (Starr 2013, 91) As suggested, sensory imagery is not only linked to embodiment and social interaction (as seen in the case of motor imagery), but it is also involved, by means of aesthetic experience, in the production of associative knowledge, triggering redefinitions and revaluations of what we feel and what we know (Starr 2013, 92).

## 4. Virginia Woolf and the Acropolis

A characteristic example of literary multisensory imagery directly related to cultural heritage can be found in Virginia Woolf's recurrent references to the Acropolis, both in her private journals as well as in her works of fiction. Woolf is among a number of prominent writers and thinkers who have traveled to and subsequently written about the renowned Ancient Greek monument throughout history, including Ernest Renan, Sigmund Freud, Henry Miller and Jacques Derrida. These remarkable accounts are inscribed in the modern literary tradition of travel narratives that can be traced back to the legacy of the 17th–18th century Grand Tour, and which has produced an extensive body of artistic and literary works around CH. Traveling, and especially the 'Eurocentric' travel tradition, in that sense, is part of an educational project always focused on a solid historical framework. Yet even as it theoretically follows the traces of cultural origins, 'voyage literature' often functions as a mirror of the self. As Dennis Porter points out in Haunted Journeys, if traveling, through the centuries, becomes a protracted act of understanding the world, travel writing is a process of understanding oneself:

*"the most interesting writers of nonfictional travel books have managed to combine explorations in the world with self-explanation. They submitted themselves to*

*the challenge of travel and, in the process, managed if not always, to make themselves over, then at least to know themselves differently". (1991, 5)*

Within this perspective, though travel writing is based on realistic conventions (i.e. explanatory narratives, explicit notes and general information), journeys to archeological remains, such as Acropolis, the Colosseum or Pompei, are often related to a bodily, shared and ineffable experience of empirical reality received through the senses.

Among the numerous allusions to the Acropolis that appear throughout Virginia Woolf's writings, the two passages included below were chosen on account of the manifest intensity of their multisensory imagery. Nonetheless, they were also deliberately procured from divergent sources by the same author so as to demonstrate how different literary perspectives might be recruited in inducing embodiment and aesthetic experience vis-a-vis CH. The first excerpt is drawn from the author's early private journals, where she recounts her personal experience of visiting the Acropolis at sunset. The second is a short passage lifted from her famous satirical novel Orlando: A Biography (Woolf 1928/2007) that alludes to the main character's transgendered metamorphosis on her perilous journey East.

As is immediately evident from the excerpts, they are both rich in visual, haptic and above all, motor imagery, owing to Woolf's perception of the Acropolis as a mighty soundless monument, in stark contrast to the noisy hustle and bustle of the Athenian streets below it. As the author writes in her early novel Jacob's Room: 'the Parthenon is really astonishing in its silent composure' (Woolf 1922/2007, 105) – an enduring impression that gives rise, all through Woolf's oeuvre, to vivid object-based, spatial, tactile and kinaesthetic images.

*"The Temple glows red; the whole west pediment seems kindled, as if for the first time, in the sunset opposite: its rays light & heat, while the other temples burn with a white radiance. No place seems more lusty & alive than this platform of ancient dead stone. The fat Maidens who bear the weight of the Erechtheum on their heads, stand smiling tranquil ease, for their burden is just meet for their strength. They glory in it; one foot just advances, their hands, one conceives, loosely curled at their sides. And the warm blue sky flows into all the crevices of the marble; yet they detach themselves, & spring in to the air, with crisp edges, unblunted, & still virile and young.*
*But it is the Parthenon that over comes you; it is so large, & so strong, & so triumphant. You feel warmed through & through, as though you walked by some genial hearth. But perhaps the most lovely picture in it – at least it is the most detachable – is that which you receive when you stand where the great Statue used to stand. She looked straight through the long doorway, made by the curved lines of the columns, & saw a long slice of Attic mountain & sky & plain, & a shinning strip of the sea. It is like a panel, let in to the Parthenon to complete its beauty. It*

*is soft, & soon grows dark, though the water still gleams; then you see that the white columns are ashy pale, & the warmth of the Parthenon ebbs from her.*
*A bell rings down below, & once more the Acropolis is left quite alone."* (Woolf 1991, 321-22)

Palpable straightaway, this description of the Acropolis at sunset generates a highly embodied experience through the multisensory combination of dramatic visual and haptic imagery of burning and of warmth: from the flaming red sunset, to the day's accumulated heat emanating from the dead, ashy pale stones, to the radiating brilliance of the mighty white-hot temple, tempered to the comforting smolder of a genial hearth. In addition, this rich multisensory depiction is further animated – making the site come alive – by vivid motor imagery of weight carried (fat & marble) with a smile, of 'one foot forward', of loosely curled hands and a sprightly leap; but also, of standing, curving, sawing and slicing until a single auditory image – the toll of a lone bell at nightfall – prevails upon, dispatches and ultimately dissipates the whole sensuous vision-occurrence.

By the same token, even though sourced from a completely different genre, the second passage from Woolf's oeuvre included below makes similar use of multisensory imagery: in this case, to capture Orlando's distant encounter with the Ancient monument. Only this time, the multisensory imagery is not employed to induce the portentous experience of a walk through: of a visit to the site of Acropolis. Instead, it is recruited to evoke the rapture of a spiritual and transformative journey.

*"There were mountains; there were valleys; there were streams. She climbed the mountains; roamed the valleys; sat on the banks of streams. She likened the hills to ramparts, to the breasts of doves, and the flanks of kine. She compared the flowers to enamel and the turf to Turkey rugs worn thin. The trees were withered hags, and sheep were grey boulders. Everything, in fact, was something else. She found the tarn on the mountain-top and almost threw herself in to seek the wisdom she thought lay hid there; and when, from the mountain-top, she beheld far off, across the Sea of Marmara the plains of Greece, and made out (her eyes were admirable) the Acropolis with a white streak or two which must, she thought, be the Parthenon, her soul expanded with her eyeballs, and she prayed she might share the majesty of the hills, know the serenity of the plains, etc. etc., as all such believers do."* (Woolf 1928/2007, 468)

Here again, Woolf enlists potent haptic and motor imagery to sustain a highly sensate voyage: a metaphorical passage, at once physical and spiritual, that requires one to climb, roam, sit and seek, at the risk of throwing one's self in, so as to behold at last, and feel one's soul and one's eyeballs swell, in prayer for sharing, knowing and believing. In this passage, Orlando's inner and outer metamorphosis is mirrored by the fact that everything is something else, and that something can be grasped to the touch: solid rugged ramparts, soft belly feathers and silky cow hide; jagged petal tooth enamel and threadbare Turkey rugs; and craggy branches and cold hefty boulders. Until, in the end, the rapid-fire succession of haptic and motor imagery gives way to a clear visual image: the Acropolis as a white streak or two in the far off distance – a revelation, born out of a transformative experience. In this last respect, it should be noted here, relevantly to the previous discussion, that the impact of this particular passage – in all of its potential for literary transportation and embodiment – is not only limited to added value around the cultural heritage monument, but also, simultaneously, induced empathy for a female transgender voice.

In contrast to the didactic pedagogical narrative content common to most DCH applications, not much can be learned from Woolf's excerpts when it comes to the Acropolis's long history and socio-cultural bearing. Nor are these two literary passages overtly entertaining – though there is certainly pleasure involved; no gaming can be fathomed from their narratives. Instead, these excerpts suggest a different type of experience in relation to CH. One that is not inherently grounded in touristic consumption – notwithstanding the first passage's plain account of a foreigner's sightseeing visit. Far from the education-entertainment-consumption paradigm at the core of a majority of DCH applications, the use of multisensory imagery in these two excerpts is rather designed to elicit an individual (re)valuation of CH through aesthetic experience. Enhancing MR digital heritage applications with gamification elements offer a unique storytelling experience which combines entertainment and education. (Ioannidis et al. 2017).

## 5. Aesthetic experience

Narrative content's ability to amplify presence through embodiment in a similar way to actual sensory input presents significant advantages for DCH applications both when it comes to the VRE and the user. As demonstrated, such narrative content can sustain accrued presence, thus fulfilling the VRE's central aim (see section 3). At the same time, it can provide effective empathy and social training for the user by recruiting literature's proven faculties to increase social abilities and alter selfhood, even in hard to reach individuals (Oatley et al. 2012). But beyond these notable effects, and more in line with the core subject of this volume, Literary-based MR Presence can also provide a distinct benefit for actual heritage sites, monuments and artefacts. Indeed, it is important to recognize that while enhanced presence, embodiment and empathy may be crucial to maintaining attention and engagement in virtual heritage applications, these factors do not ensure adequate reception of the cultural heritage itself. As Pujol and Champion observe in their study on 'cultural presence': 'ability to navigate and complete tasks in a virtual environment is no guarantee that relevant cultural learning has taken place.' (2012, 97)

Relatedly, a recent critical survey of the state of the art on Virtual Museums has found that existing DCH applications fail to live up to their radical potential owing, in large part, to their reductive pedagogical transmission of didactic content – viz. descriptive and explanatory commentary entirely disconnected from visitor expectations, personal agendas and emotional involvement (Perry et al. 2017). Perry et al., advocate instead for an emotional, participatory, interactive and social engagement of the public through the use of a collaborative and affective user-centered design methodology, aligned with the latest approaches in museum studies. Of particular relevance to the discussion at hand, the authors note that such user-centered interfaces are uniquely valuable when it comes to antiquities because: 'many heritage sites have few remnants that are either visible or relatable to the broad public. As such, they may not have enough resonance to engage visitors on their own or through standard interpretational means.' In other words, existing VM's explanatory information and empty reconstructions are unable to bring heritage sites and artefacts back to life for lack of emotionally evocative content, associated instead with storytelling media such as film and literature (Perry et al. 2017).

When compared to Perry et al.'s approach, Literary-based MR Presence shares many of the same objectives and operative modalities as their advocated model of 'emotive storytelling' – correspondingly described as an interactive, story-based (as opposed to object-based) approach involving a dramatic and affective narrative. Both, for instance, clearly aim to surpass an out-dated model of 'educative leisure', fostering instead more curiosity, attentiveness, empathy and personal transformation. Notwithstanding, Literature-based MR Presence differs on a fundamental level from Perry et al.'s affect-based approach on account of its distinct mode of reception, which is grounded in aesthetic experience.

For in effect, aesthetic experience implies, first and foremost, a process of valuation, i.e., a value judgment: it 'works to produce new value in what we see and what we feel.' (Starr 2013: 66). To be clear, value is here explicitly understood as the outcome of an evaluative judgment, and not an a priori standard, rule, criterion, norm, goal, or ideal that one might use to formulate such a judgment (Sánchez-Fernández & Iniesta-Bonillo 2007, 429). In this way, it consists of a surplus created by the user, as opposed to describing either a quality intrinsic to a cultural object, or the product of an institutional legitimation. Certainly, emotions are implicitly involved in the cognitive processes of directing attention, passing judgement (e.g., allowing individuals to make rapid decisions as to whether something is beneficial or harmful) and aesthetic experience. However, evidence indicates that the neural activity of everyday emotions differs from that associated with aesthetic emotional involvement (Starr 2013, 42). Indeed, aesthetic judgment and transportation's activation of the brain's default mode network suggests a more long-lasting and profound valuation. As follows, Literature-based MR Presence can provide a direct benefit for cultural sites, monuments and artefacts, beyond a more narrow focus on user design. It essentially opens the possibility of re-cruiting the user's attention and personal transformation towards the creation of incremental value for cultural heritage.

## 6. ViMM and the Museum 3.0

The model of valuation-response described above is entirely aligned with a specific understanding of usership emerging from new thinking at the junction of museology and social practice. Indeed, in recent years, cutting edge curatorial and educational practices have increasingly turned towards visitor-centered approaches, characterized by the unprecedented attribution of equal importance to the collection as to the public (Samis & Michaelson 2017). This has generated novel institutional methodologies implemented via physical (haptic), immersive, emotive, cognitive & co-creative, as well as meta-cognitive procedures (Samis & Michaelson 2017). Among the most radical of these visitor-centered conceptions, Arte Útil's (Useful Art) notion of the Museum 3.0 presents unique parallels with DCH applications owing to its basis in the attention economy. As philosopher Stephen Wright remarks in Toward a Lexicon of Usership, a majority of cultural institutions today have made strides to implement elements of 2.0 Culture into their operative modalities. While they have mostly retained the top-down gatekeeping mechanisms involved in the determination and dissemination of content, they have concurrently adapted their model of legitimation to incorporate visitor experience, feedback and input (2013). However, as Wright accurately observes, if museums have integrated some aspects of user-generated content, by and large the value of this contribution is far from mutualized. This is hardly surprising given that such a radical conception of usership, in contradistinction to the more commonly used term 'participation', poses a direct challenge to three fundamental aspects integral to traditional cultural institutions, namely: spectatorship, expert culture and ownership (Wright 2013).

Based on the above, uncompromising idea of usership, the Museum 3.0 thus recognizes, in essence, that user engagement generates value, and that this entitles users to share said value through a form of remunerated (though not necessarily monetized) exchange. Otherwise stated, this new vision of the museum sees usership as an engagement – or 'cognitive privilege' – that produces value as opposed to consuming it: 'usership is creation socialised, and as such engenders a surplus' (Wright 2013). While Wright explicitly affirms that he is referring to physical institutions rather than digital media and online VM, his proposed model for a Museum 3.0 essentially bridges the gap between DCH programmers' protracted efforts to emulate museum practices on the one hand, and virtual heritage applications' default mode of usership on the other. As previously stated, in contrast to traditional forms of spectatorship tantamount to a democratized elite privilege, this radical understanding of public engagement is rooted in the attention economy, which basically views human attention as a scarce commodity. In rela-tion to VE, this precious resource is the decisive condition for presence, as evidenced by the fact that it is one of the rare factors common to all varying definitions of the concept (Schuemie et al. 2001, 4). Based on this idea of the Museum 3.0, rather than monetizing attention through publicity revenues as GAFA giants have done, a ViMM 3.0 might thus alternatively channel it, at least in part, toward the creation of an ever-renewed and shared cultural heritage commons.

Literary-based MR Presence can be a useful tool towards such a project. In effect, it actualizes Roland Barthes' conception of the reader as an author in their own right, by turning the reader into a user, and valuing the contribution provided by their engagement. Instead of funnelling this value into the creation of private surplus capital, a ViMM 3.0 utilizing a Literary-based MR Presence approach might instead focus this value on personalized collective appreciation of CH. Overall, what this suggests is that in letting go of spectatorship, ownership and top-down expertise, the institutional guardians and stewards of heritage stand to gain unprecedented surplus value for the patrimony in their care, not at the outcome of endless institutional efforts at legitimization and propagation, but instead through a bottom-up model of transformative usership.

## 7. Conclusion

If heritage is to be effectively preserved, it needs to be valued by people living today. What is more, if it holds no contemporary significance for present generations, no living currency, then it consequently plays no actual role in the social fabric. The idea inherited from the Enlightenment that the public should be educated on culture and patrimony's worth through didactic means fails to recognize the process by which things come to be intimately valued. It is an out-dated, top-down, hegemonic paradigm, ill-advisedly carried over and replicated in DCH immersive applications –otherwise integrally predicated on usership – especially at a time when cultural institutions are increasingly turning towards more participatory and collaborative modes of public engagement. Virtual Museums' common pedagogical approach is thus ill-fated, and moreover generally flawed in its implementation, often necessitating recourse to gamification techniques to compensate for a lack of user enjoyment and involvement. In addition, gamified MR applications recreate a virtual environment, featuring interactive tasks to orchestrate an immersive experience (Zikas et al. 2016)

Instead of seeing public attention and engagement as wanting in information, they should alternatively be esteemed as a precious form of capital capable of generating surplus value around CH. Such valuation, however, is not automatically guaranteed at the outcome of a basic exposure to cultural heritage through digital immersion. It is contingent upon a compelling and lastingly transformative experience, which Literature-based MR Presence can provide by means of trans-portation and aesthetic experience. As shown, aesthetic experience is not only a means to greater immersion and interest, it also involves a profound evaluative judgment, being 'the result not so much of perceiving the outside world as becoming aware of our own judgment of what matters to us.' (Starr 2013, 16).

Naturally, there are trade-offs to such an approach when compared with more common design methodologies geared towards inclusionary access, didactic transmission and leisurely consumption. Aesthetic experience is highly variable from one individual, and from one cultural context, to the next (Starr 2013, 57). It is not a universal mode of reception, constrained instead to a limited audience. Certainly, personal appreciation of any literary work used in a DCH context will have an undeniable impact on the encounter, and therefore, on the resulting impression of the cultural site, monument or artefact at hand. Moreover, literature is always saddled with an issue of translation, thus implying a much more situated and audience-specific virtual museum model. While such public specificity may not be synonymous with geographical locality (e.g., people from different part of the world might have a positive aesthetic experience around the same piece of literature), it is not universal. Therefore, richness of multisensory imagery and embodiment potential should always be balanced out against public familiarity with the author, and the wide-audience appeal of the text – in contrast to more obscure niche genres – in order to maximize access and impact. Notwithstanding these acknowledged trade-offs, the fact remains that in its reconciliation of new museology and digital usership, Literature-based MR Presence holds the promise a uniquely integrated and beneficial VM model, capable of insuring personalized and ever-renewed value around cultural heritage.

15